\documentstyle[aas2pp4]{article}

\def\gyr{\hbox{$\,{\rm Gyr}$}}

\def\kpc{\hbox{$\,{\rm kpc}$}}

\def\beq{\begin{equation}}
\def\eeq{\end{equation}}
\def\beqar{\begin{eqnarray}}
\def\eeqar{\end{eqnarray}}

\def\msol{\hbox{$M_{\odot}$}}

\def\ret{\hbox{$\cal{R}$}}
\def\esn{\hbox{$\epsilon_{\rm SN}$}}

\def\asfr{\hbox{$\alpha_{\rm mrg}$}}
\def\bsfr{\hbox{$\beta_{\rm q}$}}
\def\zloss{\hbox{$\zeta_{\rm loss}$}}
\def\emix{\hbox{$\eta_{\rm mix}$}}
\def\lmrg{\hbox{$\lambda_{\rm merge}$}}
\def\lcool{\hbox{$\lambda_{\rm cool}$}}

\def\la{\mathrel{\mathpalette\fun <}}
\def\ga{\mathrel{\mathpalette\fun >}}
\def\fun#1#2{\lower3.6pt\vbox{\baselineskip0pt\lineskip.9pt
  \ialign{$\mathsurround=0pt#1\hfil##\hfil$\crcr#2\crcr\sim\crcr}}}

\begin{document}

\title{HALO WHITE DWARFS AND THE HOT INTERGALACTIC MEDIUM}

\author{Brian D. Fields and Grant J. Mathews}
\affil{Department of Physics, University of Notre Dame \\ 
Notre Dame, IN 46635 \\
{\tt bfields@cygnus.phys.nd.edu \hspace{0.75em} gmathews@bootes.phys.nd.edu}}

\and

\author{David N. Schramm} 
\affil{University of Chicago, Chicago, IL 60637, and \\
NASA/Fermilab Astrophysics Center, FNAL, Box 500, Batavia, IL, 60510 \\
{\tt dns@oddjob.uchicago.edu}}

\begin{abstract} 
We present a schematic model for 
the formation of baryonic galactic halos
and hot gas in the Local Group and the intergalactic medium.
We follow the
dynamics, chemical evolution, heat flow and gas flows of a hierarchy of
scales, including: protogalactic 
clouds, galactic halos, and the Local Group itself.
Within this hierarchy, the Galaxy is built via mergers of protogalactic
fragments. Hot
and cold gas components are distinguished, with star formation
occurring in cold molecular cloud cores, 
while stellar winds, supernovae, and mergers 
convert cold gas into a hot intercloud medium.  
We find that early bursts of star formation lead to a
large population of remnants (mostly white dwarfs), which 
would reside presently in the
halo and contribute to the dark component observed in the microlensing
experiments.  The hot, metal-rich gas from
early starbursts and merging 
evaporates from the clouds
and is eventually incorporated into the intergalactic
medium.  The model thus suggests that most microlensing objects could be
white dwarfs ($m \sim 0.5 \msol$),
which comprise a significant fraction of the halo mass.  
Furthermore, the Local Group
could have a component of metal-rich 
hot gas similar to, although less than, that
observed in larger clusters.  We discuss the
known constraints on
such a scenario and show that all local observations
can be satisfied with present data 
in this model.  The most stringent constraint
comes from the metallicity distribution in the halo.
The best-fit model has a halo that is 40\% baryonic, with
an upper limit of 77\%.
Our model predicts that the hot intragroup gas has a 
total luminosity $1.5 \times 10^{40} \ {\rm erg} \ {\rm s}^{-1}$,
and a temperature of
0.26 keV, just at the margin of detectability.
Improved X-ray data 
could provide a key constraint on 
any remnant component in the halo.

\end{abstract}


\section{Introduction}

Recently there has been renewed interest in the nature of
the dark matter in galaxy halos, motivated by the results
of microlensing experiments.  Observations toward the Magellanic
clouds (e.g., Alcock et al.\ \cite{macho95};
Aubourg et al.\ \cite{eros93}) and toward the Galactic bulge
(e.g., Udalski et al.\ \cite{ogle94}) 
have detected gravitational 
microlensing and inferred the presence of
dark, massive compact halo objects (MACHOs).  Furthermore,    
a recent binary detection in the direction of the LMC,
along with average event durations 
of about 2.5 months, may imply masses
of order $\sim 0.5 \msol$, suggestive of
white dwarfs.

At the same time, X-ray observations of clusters
(Mushotzky \cite{cluster})
and groups
(e.g., Mulchaey, Davis, Mushotzky, \& Burstein \cite{mdmb,mdmb93}; 
Pildis, Bregman, \& Evrard \cite{pbe};
Ponman, Bourner, Ebeling, \& B\"ohringer \cite{pbeb})
have discovered a large amount of hot, metal-rich gas.
Where it is observed, this gas is a substantial fraction of
the baryonic mass--it is the dominant baryonic component of
clusters, and is comparable to the galactic component in groups.
Indeed, it appears that most baryons in the universe
are in the for of this hot X-ray emitting gas.

The relatively high metallicity ($Z \sim 0.3 Z_\odot$ for
clusters, $Z \sim 0.1 Z_\odot$ for groups) of this gas is impressive
and demands that the material has undergone a significant amount
of stellar processing during an earlier  
epoch of star formation.  
Such an epoch would also produce remnants,
mostly white dwarfs. 
If this epoch is a general consequence of the formation of 
the bulge and halo of spirals like the Milky Way,
as well 
as the ellipticals of rich clusters,
it could account for the observed microlensing objects.

Although white dwarfs are attractive MACHO 
candidates, there are important constraints on such objects
and their formation
(Ryu, Olive, \& Silk \cite{ros}).
These include background light from the early evolution, 
the present luminosity of the halo, 
and the metal and helium content of the disk
and halo stars.
While these place important constraints on 
model parameters, 
they do not rule out 
a significant white dwarf halo population, as we shall show.

In modeling these galaxy aggregates and their hot gas
components, one must account for the
dependence of these systems on the
morphology of the constituent galaxies.
On the one hand, 
the hot gas in clusters
and groups
appears correlated with the luminosity of elliptical and S0 galaxies
(Arnauld et al.\ \cite{arolv}).
This suggests that the hot gas arises from the violent merging
associated with these morphological types.
On the other hand, it also seems well established
(e.g., Rich \cite{rich})
that the morphology of the bulge and halo of spiral
galaxies
is quite similar to that of ellipticals and S0's.  
This suggests that the bulge and halo of spirals 
may have experienced
a similar epoch of star formation and outflow during their
formation.  The difference in the morphologies  
may relate to
the larger angular momentum or shallower gravitational potential
of spirals (Zurek et al.\ \cite{zqs}), such that
some gas survives halo formation, and
settles afterwards into spiral arms.

With this background in mind,   
we find that a likely, and perhaps inevitable, consequence of 
the formation of the bulge and halo is the formation of 
a large remnant population
in the Galactic halo,
along with hot X-ray emitting gas in the Local Group
and intergalactic (i.e., extragroup) medium.
While our model should be widely applicable,
in this paper we concentrate on 
the Local Group.

Given the detection of dark microlensing objects, 
as well as the need for significant
amounts of dark baryonic matter {\it somewhere}
(Copi, Schramm, \& Turner \cite{cst};
Fields, Kainulainen, Olive, \& Thomas \cite{fkot})
halo white dwarfs are a very conservative candidate
(e.g., Larson \cite{lars};
Ryu, Olive, \& Silk \cite{ros}; Silk \cite{silk}).
This is particularly so since red and brown dwarfs are
apparently excluded as halo candidates (Bahcall, Flynn, Gould, 
\& Kirhakos \cite{bfgk}; Graff \& Freese \cite{gf}).
Here we make a specific though schematic model, and 
assess the plausibility of the white dwarf hypothesis.\footnote{Others
have suggested that the microlensing objects might be remnants from
an early Population III; see 
Fujimoto, Sugiyama, Iben, \& Hollowell (\cite{fsih}).}
As we will see, the model can be made to work but
not without some assumptions 
(e.g., one must alter the halo initial mass function).  
In any case, the model is eminently testable, and perhaps has
already been tested by X-ray observations in other groups.
Indeed, should one find this model and the halo white dwarf hypothesis
untenable, then it follows that the dark baryons and the
MACHOs must take an even more exotic form.

\section{Local Group Properties}
\label{sec:data}

We take the Local Group to have a 
mass $\sim (3-5) \times 10^{12} \msol$
(Fich \& Tremaine \cite{ft}).
Its luminous component is
dominated by two galaxies,
one of which (the Galaxy)
has a visible mass $\sim 7 \times 10^{10} \msol$.
The total mass of the Galaxy is uncertain and depends on the radius, i.e.,
$M_{\rm tot} \sim 5 \times 10^{11} \msol (R_{\rm halo}/50 \ {\rm kpc})$.
This implies that the dark halo has a mass within 50 kpc of
$\sim 4 \times 10^{11} \msol$, some or all of which will be 
in the form of microlensing objects and other dark baryons.  
We consider models with up to
90\% the total mass
in nonbaryonic dark matter,
at the group scale.  The baryonic fraction at the
halo scale, however, can be much less.

In many galaxy groups, hot intergalactic gas is found 
(e.g.,
Mulchaey et al.\ \cite{mdmb93,mdmb}; Pildis, Bregman, \& Evrard \cite{pbe};
Ponman, Bourner, Ebeling, \& B\"ohringer \cite{pbeb}). 
This gas is
a significant and sometimes dominant component
of the baryonic mass. The hot gas-to-galaxy mass
ratio ranges from $0.3-3$.  
The gas also contains metals, with $Z \sim (0.1-0.2) Z_\odot$.
While the ROSAT metallicity determinations are uncertain 
(Davis, Mulchaey, Mushotzky, \& Burstein \cite{dmmb}), 
recent measurements with ASCA (Fukazawa et al.\ \cite{asca})
suggest that the gas metallicities in groups could
have a larger spread than those of clusters.
In any case, the metallicity is clearly
not primordial, indicting that a significant
fraction of material has been processed in stars 
before they are incorporated into the intragroup medium.

If the gas is in
hydrostatic equilibrium with the gravitational potential of
the group, then
the temperature distribution determines the total mass.  
Furthermore, the observed temperatures 
tend to tightly cluster around $T = 1$ keV
for most groups observed thus far
(Mulchaey et al.\ \cite{mdmb}).
This implies that the 
{\it total} masses are
similar, with most observed group masses
lying in the range $(1.5 - 2.5) \times 10^{13} \msol$. 
This is a 
tighter range than the
span of {\it luminous} mass in galaxies and gas. 

However, it is not clear that all optically identified groups evidence
hot gas.  Several analyses of ROSAT observations
have found that the presence of detectable gas is strongly
correlated with the morphologies of the constituent galaxies
(e.g., Pildis, Bregman, \& Evrard \cite{pbe}).
The trend is very similar to that found in galaxy clusters:
the hot gas mass is correlated with the presence of early-type
(E and S0) galaxies.  Indeed, Mulchaey et al.\ (\cite{mdmb})
emphasize that there is at least one 
bright ($L_B \ga 5 \times 10^{10} L_\odot$)
elliptical galaxy in {\it every}
group for which hot gas has been detected.  

On the other hand, 
Ponman, Bourner, Ebeling, \& B\"ohringer (\cite{pbeb}),
also using ROSAT data,
have recently claimed the positive detection of hot gas in
spiral-dominated Hickson compact groups.  They find that the gas in these
groups has a lower temperature ($\sim 0.3$ keV), and thus, 
a lower surface brightness than that in groups with early-type galaxies.  
The discrepancies between these results
and those of previous groups attests to the 
difficulty in trying to measure or put limits on
such a relatively dim diffuse component.
Ponman et al.\ (\cite{pbeb}) 
note that differences between their results 
and others trace to details of the analysis, e.g., subtraction
of galactic and background emission.  As  
Davis, Mulchaey, Mushotzky, \& Burstein (\cite{dmmb}) point
out, these are not always straightforward 
issues.  
Hence, we regard the issue of
hot gas in spiral-dominated galaxies as presently ambiguous.

In the Local Group specifically, there has not been direct
observation of hot intergalactic gas.
However, Suto et al.\ (\cite{smio}) have argued that such
a component is allowed within current direct limits.
Indeed, they suggest that this is a source of the excess 
low-energy component
in the diffuse X-ray background.
Suto et al.\ (\cite{smio}) model
a gas distribution that could
lead to the excess radiation;  
their distribution implies a total mass in hot intergalactic gas
of about $\sim 3.5 \times 10^{11} \msol$.
Of course, the existence of hot gas in spiral-dominated
groups is a central premise in this scenario. 
Pildis \& McGaugh (\cite{pm}) show that the
observational limits on such gas (if it is similar
to other spiral-rich groups) require any 
Local Group gas to 
have too low a mass
to provide the soft X-ray background.

\section{Hierarchical Collapse Model} 
\label{sec:model}

Motivated by hierarchical clustering scenarios for
structure formation, we compute the evolution 
of a hierarchy of self-similar mass scales.
The three scales are (1) protogalactic clouds, all of which
reside in (2) galaxy halos, themselves moving within
the (3) group.  
Within a spiral protogalaxy, the clouds merge to
become ultimately
the disk and bulge.
However, we do not distinguish the disk and bulge.
Their formation does not significantly affect
the halo or group evolution once the last merging has occurred and 
the star formation (and gas outflow) in the halo has diminished.

The dynamics of the three components 
are described 
(Mathews \& Schramm \cite{ms})
as the radial evolution $R(t)$ of a spherical
overdensity, from its initial expansion (starting at
$t=10^8$ yr) through the departure from Hubble flow and
collapse to a fixed final radius.  
The halo collapse is halted at $50$ kpc.

For each component in the structure hierarchy,
we follow the evolution of matter in the form of
gas, stars and remnants, and possibly also 
non-baryonic dark matter.
We also compute the helium and metal evolution,
and follow the temperature of the hot gas.
Our model is schematic 
but contains the necessary
features for testing and constraining our basic
hypotheses.  
Its structure (summarized pictorially in
figure \ref{fig:diag}) is as follows.

\subsection{Level 1: Protogalactic Clouds}

In this simple model we assemble galaxies from
a distribution of protogalactic fragments within expanding
and contracting galactic halos.  At the level of
the protogalactic clouds, we consider the system to be composed
of three components: 
cold star-forming molecular clouds; 
heated ejecta from the 
clouds; and stars and remnants.  We assume that non-baryonic dark
matter (if there is any) is unimportant at this scale.

Let us first consider the cold star-forming gas  
and hot gas components.
These components are most dramatically
affected by star formation and merging.   Stars form
{}from the cold gas and return a fraction of their material
as hot ejecta.  This hot stellar ejecta further mixes 
with and heats the local cold gas mass into the hot component. 
In addition, material is heated during mergers 
(e.g., White et al.\ \cite{white})
as the relative kinetic energy
of the merging clouds is converted into internal energy 
of the merged system.
Also, cooling of the hot gas component returns
material back to cold star forming regions.

We write the coupled
equations for the evolution of the cold and hot gas components
for an average cloud experiencing all  
of these processes as:
\begin{eqnarray}
\label{eq:cold}
\dot m_{\rm g1}^{\rm cold}&=&-(1+ \ret \emix) \psi(t) \nonumber\\ 
&&+ \lmrg m_{\rm g1}^{\rm cold} \nonumber\\ 
&&+ \lcool m_{\rm g1}^{\rm hot} \ \ ,
\end{eqnarray}
\begin{eqnarray}
\label{eq:hot}
\dot m_{\rm g1}^{\rm hot} & = & \ret(1 + \emix) \psi(t) \nonumber\\
&&+ \lmrg m_{\rm g1}^{\rm hot} \nonumber\\ 
&&- \lcool m_{\rm g1}^{\rm hot} \nonumber\\ 
&&+ \dot{m}_{\rm in1}- \dot m_{\rm out1}^{\rm hot}   \ \ .
\end{eqnarray}
The first term on the right hand side of equations
(\ref{eq:cold}) and (\ref{eq:hot})
describes the formation of stars and ejecta.
\ret\ is the usual (e.g., Tinsley \cite{tins}) 
returned fraction of material from stars
in the instantaneous recycling approximation.
It depends on the initial mass function.  
The hot ejecta is assumed to sweep up and mix with the cold
cloud material; the resulting mixture contributes
to the hot gas component.  We therefore include
an additional factor of \emix\ describes the number
of equal masses of local interstellar material
that is heated into the hot component along with the stellar ejecta. 
The rate of incorporation of
cold gas into new stars is $\psi(t)$.   

The second term on the
right hand side in equations (\ref{eq:cold}) and (\ref{eq:hot})
describes the effects of mergers.  Here, \lmrg\
is the merger rate per cloud.  On average, during each merger,
the cloud mass will double.  The third term describes the cooling from
hot to cold gas, with the cooling rate 
per particle given by $\lcool$.
The last terms in Eq. (\ref{eq:hot}) accounts
for infall and outflow of gas.  
The infall term $\dot{m}_{\rm in}$ gives the rate at which hot halo gas
is incorporated int to the 
hot gas component of the clouds.  
The outflow rate $\dot{m}_{\rm out1}^{\rm hot}$
we attribute to the evaporation of hot gas.  
The physics behind each of these terms is discussed in
\S \ref{sec:phys}.

For our purposes, we take the cold star forming gas to be truly cold
gas with $T \la 100$ K.  The hot gas component is approximated
by assuming that all gas at higher temperatures is
isothermal and homogeneous
within each of the levels of structure.

If we combine equations (\ref{eq:cold}) and (\ref{eq:hot}) 
we can recover the
familiar instantaneous recycling equation for one gas
component; 
\begin{equation}
\dot m_{\rm g1}^{\rm cold} + \dot m_{\rm g1}^{\rm hot} =
-(1-\ret)\psi(t) + \dot m_{\rm in}^{\rm eff} - \dot m_{\rm out} \ \ ,
\end{equation}
where the effective infall rate, due to mergers 
as well as the influx of halo gas, is 
$\dot m_{\rm in}^{\rm eff} =
(m_{\rm g1}^{\rm cold} + m_{\rm g1}^{\rm hot}) 
\lmrg+\dot{m}_{\rm in1}$.
The outflow rate is just 
$\dot m_{\rm out} = \dot m_{\rm out1}^{\rm hot}$.

As in MS93 we  assume that
protogalactic mergers disperse the stars and remnants into the halo.
The evolution equation for the mass $m_{r1}$ 
in stars and remnants
remaining in a cloud
becomes:
\begin{equation}
\label{eq:rem1}
\dot m_{\rm r1} =  (1-\ret)\psi(t) 
             + (1-\kappa) \lmrg m_{\rm r1} \ \ .
\end{equation}
Here $\kappa$ is a measure of
the efficiency for the mergers to disperse stars and
remnants.
Specifically, the average fraction $f$ of stars and remnants
born at time $t_0$ which survive merging up to time $t$ is 
\beq
f = e^{- \kappa \int_{t_0}^t \ dt^\prime \lmrg}
\eeq
We will take $\kappa=1$.

Since stars return all of their
ejecta into the hot component, the metallicity in
the cold star forming gas is only indirectly enriched by cooling from
the hot component.  Thus, from Eq.\ (\ref{eq:cold}) we write for
the evolution of the total mass in metals in the cold star forming gas
$(Z_1^{\rm cold} m_{\rm g1}^{\rm cold})$:\footnote{We explicitly show only
the metal evolution here and below, but from this
the expressions for the helium evolution follows trivially.}
\begin{eqnarray}
\frac{d (Z_1^{\rm cold} m_{\rm g1}^{\rm cold})}{dt}
&=& Z_1^{\rm hot} \lcool m_{\rm g1}^{\rm hot} \nonumber  \\ 
&& - Z_1^{\rm cold} \biggl[ (1 + \ret \emix) \psi(t) \nonumber  \\
& &  + m_{\rm g1}^{\rm cold} \lmrg \biggr] \, \, .
\end{eqnarray}
This reduces to a simple evolution equation for 
the metallicity of the cold gas,
\begin{equation}
\dot Z_1^{\rm cold} = \left( Z_1^{\rm hot} - Z_1^{\rm cold} \right) 
\lcool {m_{\rm g1}^{\rm hot} \over m_{\rm g1}^{\rm cold}} \ \ .
\label{eq:zc1}
\end{equation}
Clearly, the equilibrium metallicity of the cold star-forming
component is equal to the hot component metallicity.
This equilibrium is, however, only achieved after  
a time given by the cooling
time scale, $\lcool^{-1}$
times the ratio of cold to hot gas masses.

By an analogous process to the derivation of Eq. (\ref{eq:zc1}),
the evolution of metallicity in the
hot gas component can be written
\beqar
\dot{Z}_1^{\rm hot} 
  & = & \left[y_Z + (1+\emix) \ret (Z_1^{\rm cold} - Z_1^{\rm hot}) \right]
        \frac{\psi(t)}{m_{g1}^{\rm hot}}  \nonumber \\
  & & + (Z_{\rm g2}^{\rm hot} - Z_{\rm g1}^{\rm hot}) 
        \frac{\dot{m}_{\rm in1}}{m_{\rm g1}^{\rm hot}}
\eeqar
where $y_Z$ is the mass fraction of newly synthesized material in the
ejecta.
As usual, $y_Z$ depends on both the initial mass function,
and on the stellar nucleosynthesis yields as a function of
mass (and metallicity).  

By a similar derivation, the temperature of hot gas in the halo is 
determined from an
energy balance equation.  This leads to an evolution of internal 
energy per unit mass
$\epsilon_1$ in the hot gas,
\begin{eqnarray}
\label{eq:nrg1}
\dot \epsilon_1&=&\biggl[\esn - \epsilon_1(1 +\emix)\biggr]
    \frac{\ret \psi(t)}{m_{\rm g1}^{\rm hot}} \\
&&\nonumber 
   + \lmrg \epsilon_{merge} \\
&&\nonumber 
   - \left(\epsilon_1^{\rm out} - \epsilon_1\right)
     {\dot m_{\rm out}^{\rm hot} \over  m_{\rm g1}^{\rm hot}} \\
&&\nonumber
   + \left( \epsilon_2 - \epsilon_1 \right) 
     \frac{\dot{m}_{\rm in1}}{m_{\rm g1}^{\rm hot}} \\
&&\nonumber 
   - 2 \frac{\dot R_1}{R_1} \epsilon_1 \ \ , 
\end{eqnarray}
where \esn\ in the average energy per unit mass
in all stellar ejecta averaged over an appropriate initial
mass function.  This term is dominated by supernovae
whose energy release is  $E_{\rm ej} \sim 10^{51}$ erg  per supernova.  
Averaging over the initial mass function $\phi(m)$
then gives 
\beq
\esn  =  \frac{ \int \ dm \ \phi(m) \ E_{\rm ej}(m) }
              { \int \ dm \ \phi(m) \ m_{\rm ej} }
\eeq
which is independent of the normalization convention of $\phi$.
For a typical initial mass function, this gives values of order
$\esn \sim 10^{49}$ erg M$_\odot^{-1}$.

The quantity $\epsilon_1^{\rm out}$ is the average energy per unit mass for the
material exiting the cloud. 
For an ideal gas, the temperature is simply related to $\epsilon_1$ by,
$T_1= (2 \epsilon_1 \mu / 3 k N_A)$, 
where $\mu$ is the mean molecular weight of
the gas and $N_A$ is Avagadro's number.  

The last term in Eq.\ (\ref{eq:nrg1})
accounts for the $p \; dV$ work done as
the cloud expands or contracts; the radial dependence is given
by the collapse of a spherical overdensity.  
The cloud radius $R_1$ is the 
average tidal radius
\begin{equation}
R_{\rm 1} = \biggl(\frac{m_1}{m_2}\biggr)^{1/3}R_2~~,
\label{rtidal}
\end{equation}
where $m_1$ is the total average cloud mass
\begin{equation}
m_1 = m_{\rm g1}^{\rm cold} + m_{\rm g1}^{\rm hot} + m_{r1}~~,
\end{equation}
and $m_2$ is the average halo mass defined 
below (Eq.\ \ref{eq:mhalo}).
We assume that the relative momentum 
of the merger goes into heating the cloud gas.  
Thus, we have 
$\epsilon_{\rm merge} = E_{\rm merge}/M_{\rm merge} =  v_{\rm rel}^2/2$.

\subsection{Level 2: Galactic Halos}
\label{sec:cloud}

For our purposes, galactic halos are treated as a homogeneous assembly
of protogalactic clouds and hot gas, possibly having a component of
nonbaryonic dark matter.  The total mass
of the galactic halo is then
\begin{equation}
\label{eq:mhalo}
m_2 = n_{\rm c} m_1 + m_{\rm g2} + m_{r2} + m_2^{\rm NB} \ \ ,
\end{equation}
where $n_{\rm c}$ is the number of protogalactic clouds, $m_{\rm g2}$
is the mass of hot gas which has exited the clouds to reside in the
halo, and $m_{r2}$ is the mass of stars and remnants which have
been dispersed from clouds into the halo.  
The mass of the nonbaryonic component, if present, is $m_2^{\rm NB}$. 

The equation describing
the evolution of the (hot) gas component 
in the halo is:
\beqar
\dot m_{\rm g2}
  & = & n_{\rm c} \left( \dot{m}_{\rm out1} - \dot{m}_{\rm in1} \right) \\
\nonumber &&
  - \dot m_{\rm out2} + \dot m_{\rm in2} \ \ ,
\eeqar
 where $\dot m_{\rm out2}$ is the rate at which the hot gas is
ejected from the halos as described below and $\dot m_{\rm in3}$
is the possible inflow of gas from the intragroup medium.

The rate at which stars and remnants are injected into 
the halo from mergers can be 
inferred from Eq.\ (\ref{eq:rem1}),
\begin{equation}
\dot m_{\rm r2} = \kappa n_{\rm c} m_{\rm r1} \lmrg~~. 
\end{equation}

The equation governing the evolution of metallicity in the halos
will be
\beqar
\dot{Z}_2 & = &
  (Z_1^{\rm hot} - Z_2) {n_{\rm c} \dot m_{\rm out1}\over m_{\rm g2}} \\
\nonumber & &
 + (Z_3 - Z_2) 
  \frac{\dot{m}_{\rm in2}}{m_{\rm g2}}  \ \ , 
\eeqar
where $n_{\rm c}$ is the number of cold protogalactic clouds in the halo.
Similarly the internal energy per unit mass, $\epsilon_2$,
of gas in the halo is determined from an
energy balance equation:
\begin{eqnarray}
\dot{\epsilon}_2 & = & 
  (\epsilon_{1}^{\rm out} - \epsilon_2) 
  {n_{\rm c} \dot m_{out1}^{\rm hot} \over m_{\rm g2}}  \\
\nonumber &&
  - \biggl(\epsilon_2^{\rm out} - \epsilon_2\biggr)
    \frac{\dot m_{out2}}{m_{\rm g2}} \\
\nonumber &&
  + \left( \epsilon_3 - \epsilon_2 \right)
    \frac{\dot{m}_{\rm in2}}{m_{\rm g2}} \\
\nonumber &&
  - \lcool \epsilon_2
  - 2 \frac{\dot R_2}{R_2} \epsilon_2 ~~ . 
\end{eqnarray}

The number of clouds decreases exponentially 
with the number of mergers:
\begin{equation}
\label{eq:ndot}
\dot{n}_c = - \lmrg n_{\rm c} \ \ . 
\end{equation}

\subsection{Level 3: Hot Intra-Group Medium}

The equations governing the evolution of the 
intragroup medium are analogous
to those for the galactic halos, but at the level of the group
nonbaryonic dark matter may be a dominant
contributor.  Thus, for the total group mass we write,
\begin{equation}
m_3 = n_{\rm h} m_2 + m_{\rm g3} + m_3^{\rm NB} \ \ , 
\end{equation}
where $n_{\rm h}$ is the number of galactic halos in the group,
$m_{\rm g3}$ is the mass of the hot X-ray intragroup medium, 
and $m_3^{\rm NB}$ is
the contribution from nonbaryonic dark matter.

The evolution equation for the intragroup medium is then,
\begin{equation}
\dot m_{\rm g3} = n_{\rm h} (\dot{m}_{2out} - \dot{m}_{\rm in2})
- \dot m_{3out} ~~, 
\end{equation}
and 
$\dot m_{3out}$ is the rate at which the gas is
lost from the group.

The equation governing the evolution of metallicity in the 
intragroup medium
is then
\begin{equation}
\dot Z_3 = (Z_2 - Z_3) {n_{\rm h} \dot m_{2out}
\over m_{\rm g3}} ~~, 
\end{equation}
and the energy balance equation is
\begin{eqnarray}
\label{eq:nrg3}
\dot \epsilon_3&=&(\epsilon_{2}^{\rm out} - \epsilon_3)
 {n_{\rm h} \dot m_{\rm out2}
 \over m_{\rm g3}}\nonumber\\
&&- \left(\epsilon_3^{\rm out} - \epsilon_3 \right)
    \frac{\dot m_{out3}}{ m_{\rm g3}} \\
&&\nonumber 
  - \lcool \epsilon_3
  - 2 \frac{\dot{R}_3}{R_3} \epsilon_3 \ \ , 
\end{eqnarray}
{}from which temperature of the intragroup medium 
can be inferred.

\subsection{Cooling, Star Formation, Mergers, and Mass Loss}
\label{sec:phys}

The evolution is given by equations
(\ref{eq:cold}--\ref{eq:nrg3}). What remains is to specify
the various input quantities, i.e., the
rates for cooling, star formation, merging, and mass loss.

The metallicity-dependent cooling rate \lcool\
is derived from the calculations of
B\"ohringer \& Hensler (\cite{bh}).  For the temperatures
appropriate for the hot gas, the cooling is dominated by 
brehmsstrahlung emission from electrons.
These losses are written in terms of
the cooling function $\Lambda$, which
gives the energy loss rate per unit density.
Given the number density at a given scale $n_i$, one
may then compute the energy loss rate per particle:
$\dot{E_i} = n_i \Lambda$, and then the cooling
rate $\lcool \equiv \dot{E}/E$.

We allow star formation to be induced both by mergers and intrinsic
star formation processes within the clouds (MS93).  In MS93 it was
shown that intrinsic star formation is more important during the
subsequent evolution as material settles into a disk. 
To describe
the merger-induced star formation for all of the merging substructures
within a collapsing
halo, MS93
proposed a schematic model of colliding virialized protogalactic
clouds within an expanding and contracting halo. 
We adopt that formulation here.   

Specifically, we presume that the 
fragments virialize as they form.
The collision rate 
per cloud within this virialized velocity distribution can be written,  
\begin{equation}
\lmrg = (n_{\rm c}-1)\sigma v/V\ \   , 
\end{equation}
where $n$ is the number of protogalactic clouds within a volume $V$, 
$\sigma$ is an average collision cross section, 
and $v$ is the virial velocity ($v \sim [0.4GM_{\rm h}/R_{\rm h}]^{1/2}$), 
where $M_{\rm h}$ is 
the total gravitational mass of a galactic halo, 
and $R_{\rm h}$ is the radius of the galactic halos,
approximated as a collapsing spherical overdensities.  
We define
halos as those regions which evolve to become independent gravitationally
bound ensembles of gas and stars at the present time.
All halos themselves are viewed as the result of merging internal
structure. Hence we only describe merging within the halos and not
merging between halos.  This is largely a matter of semantics.

The number of protogalactic clouds decreases exponentially with the
integral of the merger rate (Eq.\ \ref{eq:ndot}),
\begin{equation}
n_{\rm c}(t) = n_{\rm c}(0) \ e^{-\int_0^t \lmrg dt'}\ \ , 
\end{equation}
where the initial number of clouds, $n_{\rm c}(0)$, is given by the ratio of
the total initial halo baryonic mass to the initial cloud mass.  Thus,
$n_{\rm c}(0) \sim 10^6$ for this schematic model.  For the merger cross
section $\sigma$ we use $\sigma = \pi R_{\rm t}^2$,
with a tidal radius $R_{\rm t} = 2 R_1$ to include
gravitational effects (Binney \& Tremaine \cite{bt87}).

Near the end of the collapse of the halos, 
the combined effects of
conservation of angular momentum and heating will dissipate the
radial motion.  In our schematic model, we approximate these effects by
halting the halo collapse at a size of a typical present dark-matter
halo, ($R_2 \sim 50 \kpc$).  The cloud radii are fixed by Eq.\
(\ref{rtidal}), and the group collapse is halted when the radius
reaches the present 
size of the Local Group, $R_3 \sim 700$ kpc.
 
The merger-induced stellar birth rate thus  is taken 
as proportional to the mass of gas mass participating 
in mergers per unit time.
The intrinsic, quiescent star formation rate is
taken as $\propto \rho_{\rm cold}^n m_{\rm gas1}^{\rm cold}$, where
$n=1/2$ in our case.  
The total star formation rate is the sum of the two 
terms:
\begin{equation}
\label{eq:sfr}
\psi(t) = \left( \asfr \lmrg + \beta_Q \rho_{\rm cold}^{1/2} 
      \right) m_{\rm g1}^{\rm cold} 
~~, 
\end{equation}
Note that we have parameterized the strength of merger-induced
star formations in terms of the dimensionless
efficiency \asfr.  Typical numbers for \asfr\
are $\sim 1\% $ (MS93).  The coefficient of the
quiescent piece is taken as $\beta_Q = \bsfr \Lambda_0$,
where 
$\Lambda_0 = 1.7 \times 10^{-4} / 
  (\msol \ \kpc^{-3})^{-1/2} \ {\rm Gyr}^{-1}$. 
Typical values of the dimensionless scaling \bsfr\ are
$1-5$ (MS93).

Finally, we must specify the rate at which hot gas is
ejected from, or falls into, the
clouds, halos, and the group.  
We first calculate the outflow due to
evaporative mass loss.
We assume that the
hot gas is distributed homogeneously within an object.  The rate of loss
of hot gas is then just given by the fraction of a Maxwellian
thermal distribution of
velocities in excess of the escape velocity at the surface of the
structures at all three levels.  Thus, we write,
\beqar
\nonumber
\dot m_{{\rm g}i}^{\rm hot} & = & 3 \zloss \frac{\langle v(>v_e) \rangle}{R_i} 
    m_{{\rm g}i}^{\rm hot} \\
& \equiv & \lambda_{\rm out} m_{{\rm g}i}^{\rm hot}  
\eeqar
Where, $i$ denotes clouds, halos, or the group,
and $\zloss \le 1$ is a dimensionless
scale factor depending on the geometry
of the cloud; $\zloss = 1$ for a sphere.
The factor $\langle v(>v_{\rm esc}) \rangle$ 
is the average velocity of all particles above
the escape velocity $v_{\rm esc}^2 = 2GM_i/R_i$:
\beqar
\langle v(>v_{\rm esc}) \rangle & = & 
  \frac{\int_{v_{\rm esc}}^\infty dv \ v^3 \; f_{MB}(v)}
       {\int_0^\infty dv \ v^2 \; f_{MB}(v)}
\nonumber \\
  & = & \frac{1}{4 \sqrt{\pi}} \ (1+x_i)e^{-x_i} \  v_T 
\eeqar
where $f_{MB}$ is the Maxwellian distribution.
The average thermal velocity is $v_T^2 = 2kT_i/\mu m_p$, and
$\mu m_p$ is the average mass of a gas particle.
$x_i = GM_i/R_i k T_i$ is the ratio of gravitational binding energy
to thermal energy, and is small typically.

Similarly, the average energy loss per unit ejected mass of material is
\beqar
\nonumber
\epsilon_{i}^{\rm out} & = &
  \frac{\langle \epsilon v (>v_{\rm esc}) \rangle}
  {\langle v(>v_{\rm esc}) \rangle} \\
& = & \frac{1+x_i+x_i^2/2}{1+x_i} \ \frac{2kT_i}{\mu m_{\rm p}} 
\eeqar
which for small $x_i$ gives 
$\epsilon_{i}^{\rm out} \simeq 2kT_i/\mu m_{\rm p} = 4\epsilon_i/3$.

Using similar derivations, we have computed inflow 
of halo (group) gas into the clouds (halos).
We note that this inflow has two components.  First,
some material is accreted simply due to the motion of
the halo or cloud through the surrounding medium--the finite
size of these substructures will lead to the capture
gas that falls within their geometric cross section
$\sigma = \pi R_i^2$.  Second, the hot gas particles
surrounding the
substructures 
will accrete due to their random thermal motion.
We assume that these particles quickly equilibrate
with the ambient gas.
Together these two effects lead to an infall term of the form
\beq
\dot{m}_{{\rm in}i}^{\rm hot} 
   =  \frac{4}{3} \ \zloss \ \left( \frac{R_{i}}{R_{i+1}}\right) \ 
      \frac{m_{{\rm g}i+1}^{\rm hot}
      \left( v^{T}_{i+1} + v_{\rm c} \right)}{R_{i+1}} \ ~~~~,
\eeq
with $v_{\rm c}^2 \equiv Gm_{i+1}/R_{i+1}$ 
the cloud (halo) circular velocity.

\subsection{Initial Mass Function and Stellar Yields}
\label{sec:imf}

The initial mass function (IMF) is important for our scenario
in several respects.  As we have noted, it 
affects the parameters of our model, namely the
returned fraction \ret, yields $y_Y$ and $y_Z$, and 
specific energy injection \esn.  Moreover, the crucial 
impact of the IMF for our scenario is that it
determines the ratio of low mass stars to high mass stars
in the halo.  Low mass stars
($m \la 0.9 \msol$ for halo metallicities)
will still be burning today.   While these stars are faint, 
they would be detectable, and 
heavily favored by an IMF 
similar to that of present disk stars.
As a result, such stars would today outnumber by far the halo
remnants, and their
net contribution to the halo luminosity would be large
(Ryu, Olive, \& Silk \cite{ros}).
Thus, remnants cannot be a significant component of the halo
if the halo IMF is similar to that inferred for disk stars. 
To allow for a significant population
of halo white dwarfs,
they must have been formed from an IMF which strongly
favored the formation of intermediate- to high-mass stars
over low-mass stars
(Ryu, Olive, \& Silk \cite{ros}).

Given the different physical conditions 
during the formation of the halo
(e.g., frequent mergers,
higher temperatures,
lower metallicities), it is at least plausible
that the star forming process then was different than
that of disk stars.  
Indeed, there are arguments for an early IMF that is
skewed towards higher masses
(see Silk \cite{silk}; 
Bond, Arnett, \& Carr \cite{bac};
Adams \& Fatuzzo \cite{af}).

Therefore, we have chosen the simple log-normal IMF
parameterization, as suggested by
both observational (Miller \& Scalo \cite{milscal}) and
theoretical arguments (Adams \& Fatuzzo \cite{af} and refs therein):
$\ln \phi(\ln m) = \ln \phi_0 - [ \ln^2(m/m_c)/2\sigma^2 ]$.
We investigate the effect of variations of the centroid
mass $m_c$ and the dimensionless width $\sigma$.
We take the IMF bounds to be
(0.1,100)\msol, with a black hole cutoff at 18\msol (Brown
\& Bethe \cite{bb}).  In fact, our
results for metal yields are insensitive to these 
limits.  With these parameters, we obtain a returned fraction
$\ret = 0.375$.

The stellar yields are taken from Maeder (\cite{mae}).  We follow both
helium and metallicity $Z$.  These abundances are particularly
powerful constraints when used together.  The helium yields are
mainly from intermediate mass stars ($m \la 8 \msol$), while
the metal yields come primarily from high mass stars.  Thus,
the IMF must 
strike a balance between the two mass ranges 
to avoid an inappropriate ratio of helium to
metals.  With our IMF, we find $y_Y = 0.0139$ and
$y_Z = 0.0354$.

\section{Results}
\label{sec:results}

In this hierarchical merging picture 
we take the clouds to be the initial building blocks of all structure.
Thus, while we assume the clouds to be comprised 
initially of hot and cold gas, 
we assume baryons in the halos and the Group to be 
initially within clouds only.   Thus, by definition the initial 
gas in the halos and group is zero.  The initial metallicity
is zero, while we take a primordial helium abundance of
$Y = 0.235$.

We have run the model for plausible ranges of 
the input parameters.
The two models we present here are: (1) a ``best-fit'' model
which has appropriate final masses and metallicities,
while also optimally satisfying other constraints 
(\S\S \ref{sect:budget}--\ref{sect:lume});
and (2) a ``maximum remnant'' model which has the highest possible
halo remnant mass without violating the constraints.
Parameters of these models are summarized in Table 1.
Unless otherwise noted, numerical results will be given for the
best-fit model.

For all models, the qualitative results are as follows.
There is generally a high initial merger rate, which lasts for
$\la 1$ Gyr.  This reduces
the number of protogalactic clouds from $10^6$ to about 100
while producing a burst of star (and hot gas) formation.  
The star formation rate and cloud number evolution appear
in Figure \ref{fig:sfr}.
Some of the stars are ejected into the halo by the mergers,
and on a longer timescale, a wind is ejected---first from clouds,
then from halo, and ultimately from the Group.  
When the halo collapses, after $5 \gyr$
(Mathews \& Schramm \cite{ms}),
there is an additional burst of merging and star formation. 
The remaining clouds coalesce into 
what will eventually become the galactic bulge and spiral arms.
With the collapse there is also heating and a reinvigorated 
wind.  Subsequently, all
remaining (hot, metal-rich) halo gas 
is ejected into the intergalactic medium.
The evolution of halo and group masses 
is given in Figure \ref{fig:masses}.

\subsection{Disk and Bulge Formation}

The clouds begin with the Jeans mass at recombination,
$10^6 \msol$.  They merge to form the proto-disk+bulge, with a mass
$8.1 \times 10^{10} \msol$.  It is encouraging that,
at the end of the halo collapse,
most material in the proto-disk
is in cold gas 
which will form disk stars.
The evolution of metallicities is given
in Figure \ref{fig:mets}.
As seen in the figure, the metallicities 
grow rapidly in the initial burst, then remain
fairly constant until the halo collapse at 5 Gyr.
Then the metals rise again as the halo collapses
and the disk and bulge are formed.  Hence, the
halo evolution provides an initial metallicity
for the gas of the protodisk.  This corresponds to
an ``initial enrichment'' of the disk material, and
so avoids overproduction of metal poor disk stars 
(i.e., the disk G-dwarf problem).

\subsection{Halo Formation}

In the best-fit model, portrayed in 
Fig.\ \ref{fig:masses}a,
the galaxy (=clouds+halo) begins with a mass of 
$1.35 \times 10^{12} \msol$, of which
$1.1 \times 10^{12} \msol$ 
is baryonic (all in the clouds).  
Even higher fractions are allowed, as discussed
below (\S \ref{sect:machofrac}).
The final galaxy mass is $5.0 \times 10^{11} \msol$,
of which $2.5 \times 10^{11} \msol$ is baryonic;
thus 50\% of the total Galactic mass is in baryons.
Some of these baryons constitute the disk+bulge,
as described above.  The baryonic mass of the halo remnants alone
(excluding the disk+bulge)
is $1.7 \times 10^{11} \msol$, and the halo nonbaryonic mass
accounts for $2.5 \times 10^{11} \msol$.  
Hence, 40\% of the dark halo is in remnants,
consistent with the microlensing observations.

When combined with mass loss from the
other galaxy in the Local Group, 
there is a total loss of
$1.7 \times 10^{12} \msol$ of gas from both halos
into the intragroup medium.
The galactic wind 
is thus very efficient, as it must be to remove the ejecta
that accompanies the remnant production.
The remnants themselves are mostly 
white dwarfs, with 12\% neutron stars.

\subsection{Group Evolution}

In the best-fit model, the Local Group
begins (Fig.\ \ref{fig:masses}b)with a mass of $5.6 \times 10^{12} \msol$, 
and a baryonic mass of $2.2 \times 10^{12} \msol$.  
At the end of the simulation
the group has a total
mass $4.3 \times 10^{12} \msol$, 17\% of which is baryonic.
Of the baryonic group mass, $2.9 \times 10^{11} \msol$ resides in hot gas.
The intragroup gas temperature is 0.26 keV, just at the limit of ROSAT
sensitivity.
The group as a whole loses $1.4 \times 10^{12} \msol$ 
of gas to the intergalactic medium; 
thus about 64\% of the initial baryonic mass in the group
is ejected later into intergalactic space.  

It is remarkable that the remnants in the halos, 
turn out
to represent a small fraction of 
the initial baryonic matter.  The 
galactic winds required to remove the stellar ejecta co-produced with
the white dwarfs
prove strong enough to remove most gas from the group itself.
This amount of hot 
(ionized) material
is consistent with Gunn-Peterson limits on the intergalactic medium
{\it if it does not cool further} (\S \ref{sec:xray}). 

Thus, in the best-fit model we find the
dark halo to be 40\% microlensing objects.
These take the form of stellar
remnants, 88\% of which are white dwarfs, and the
rest neutron stars (and perhaps black holes).
In this model the copious production of
hot intragroup and intergalactic gas is a
natural consequence of white dwarf-dominated halos.
We produce a present mass of
intragroup gas of $2.9 \times 10^{11} \msol$. 
This corresponds to about 37\% of the
baryonic mass of the Local Group.  
This mass is
encouragingly near the value ($3.5 \times 10^{11} \msol$)
implied
by the Suto et al.\ (\cite{smio})
model for the excess diffuse X-ray background (\S \ref{sec:data}).
(But recall the observational controversy \S \ref{sec:data}.)

The X-ray luminosity of the halo is 
$1.5 \times 10^{40} \ {\rm erg} \ {\rm s}^{-1}$.
Let us assume for simplicity that
the earth is at the center of this emission,
and that the gas
extends homogeneously to the edge of the
Local group ($R_{3} = 700$ kpc in this model). 
This leads to a total diffuse background of about 
$4 \times 10^{-2} \ {\rm erg} \ {\rm s}^{-1} \ 
{\rm cm}^{-2} \ {\rm sr}^{-1}$.  This is, at its
peak energy, about 3 orders of magnitude
below the observed
diffuse soft X-ray background.  We therefore find the diffuse Local Group
radiation to be negligible.  The contribution is thus
much lower than that 
suggested by Suto et al.\ (\cite{smio}), who
postulate a similar (but larger) 
intragroup gas mass, but posit a temperature 
of 1 keV as opposed to 0.26 keV in this model.

\subsection{Mass budget}
\label{sect:budget}

The most basic constraint on our model is the
requirement that it reproduce the observed mass 
and metal budget 
of the Local Group.  
The final masses determine the initial masses,
but the metallicity values constrain how to get these.
Consider the total metal mass production in our model.
The total initial
baryonic mass is $2.2 \times 10^{12}$.
Of this, the mass processed into stars and remnants
is $M_{rem}= 4.95 \times 10^{11} \msol$.
The total mass into ejecta is just
$\ret/(1-\ret) / M_{rem}= 3.0 \times 10^{11} \msol$.
The total metal mass produced is 
$y_Z M_{\rm ej} = 1.1 \times 10^{10} \msol,$ 
and the total new helium mass is 
$y_Y M_{\rm ej} =  2.8 \times 10^{10} \msol.$ 

Now, the average {\it new} metals from a given star
represents a high fraction of that star's ejecta:
$y_Z = 0.014 \sim 2/3 \ Z_\odot$.  
However, this metal-rich ejecta is diluted in several ways.
First, it is mixed with $\eta$ times its mass as it leaves
the cold star-forming regions.  Then it is mixed into
the hot cloud gas, which quickly evaporates into the halo
and on to the intragroup medium.  
Much gas is even evaporated from the Local Group 
itself, becoming part of an intergalactic (extragroup) medium.  
Thus, the final mean metallicity must be averaged over all
of the baryons (including the ejecta from the Local Groups).
One finds that the global
average mass fraction is just $0.0049 \simeq 0.25 Z_\odot$, an
acceptably small value.  

The key point here is that star formation does 
not occur in all 
regions containing gas (as is often assumed in simple chemical
evolution models)
but only cold molecular cloud cores.
Consequently, the stars represent a small 
fraction of the total baryonic mass.  
Furthermore, whereas the stars
remain in the galaxies, the gas is dispersed at all levels.
The halo can thus contain many remnants but
not a large amount of metals.

\subsection{Nucleosynthesis}

Nucleosynthesis provides an important additional constraint
on this scenario
(Ryu, Olive, \& Silk \cite{ros};
Charlot \& Silk \cite{cs};
Hegyi \& Olive \cite{ho}).  The 
burst of star formation which produces the white dwarfs
must not overproduce metals and helium.
As noted above, the metallicities
at the different scales are reasonable.
Furthermore, we may test not only the average
metallicities, but also the halo metallicity distribution.
Figure \ref{fig:hmdis} compares our calculated 
halo stellar metallicity distribution 
with the observed globular cluster distribution.
The agreement is good for the most metal-poor ($[Z] < -1$),
and thus the oldest, portion of the population.

The halo metallicity distribution strongly constrains
the total halo remnant mass in our model.  
The position of the peak constrains the net
star formation rate amplitude.  If the star formation rate is
too large, the peak is shifted too high.  Thus, the need
to reproduce the peak at $[Z]  \sim -1.5$ limits
the mass processed into stars, and so leaves
fewer stars available for the halo.  
From the $\chi^2$ of a fit to the
halo metallicity distribution we estimate a 
$2 \sigma$ upper limit of
a 77\% remnant contribution
to the halo.
Consequently, it seems likely that the
halo contains at least some
non-baryonic dark matter.
In addition, the shape of the distribution is
largely sensitive
to the relative star formation rate, and to the
degree of mixing between the stellar ejecta with the
cold gas.  

The IMF shape is of course important in
determining the metallicities.  
However, contrary to Ryu, Olive, \& Silk, we find no
significant constraint on the IMF upper limit,
essentially
because the ejection of gas removes some of the metals produced
by high-mass stars.
This is even more so if there is a cutoff for black hole formation as low
as 18 \msol\ (Bethe \& Brown \cite{bb}).  This is

\subsection{Halo Luminosity}
\label{sect:lume}

Several sources of luminosity also provide important constraints.

(1) Low mass stars in the halo are long lived and can lead to
an unacceptably high halo mass-to-light ratio
(Ryu, Olive \& Silk \cite{ros}; Richstone, Gould,
Guhathakurta, \& Flynn \cite{rggf}); however, in our model, the
mass to light ratio is comfortably above these authors 
limit $M/L_B > 500$ for $R_{\rm halo} = 50$ kpc.
We determine the luminosity contributed
by all stars
still burning in the halo by integration of 
stellar luminosities, the past star formation rate, and initial mass function,
without the instantaneous recycling approximation.  
We find a mass to total luminosity ratio of $M/L_{\rm tot} = 760$, 
$M/L_{V} = 1300$, and $M/L_{B} = 3500$.

(2) Light from the bursts of star formation can lead to
a large diffuse background (Charlot \& Silk \cite{cs};
Zepf \& Silk \cite{zs}).  Using a population synthesis model
we have preliminarily computed this background assuming the starlight
to be unscattered after its emission.  
These will be discussed in Mathews et al.\ (\cite{paperii}), but
results are consistent with observed constraints.
We exceed the
Charlot \& Silk (\cite{cs}) limit of a 10\% remnant halo
because most of the elements are formed at very large
redshift ($z \sim 10)$.

(3) The halo white dwarfs will contribute to the 
white dwarf luminosity function
(e.g., Adams \& Laughlin \cite{al};
Chabrier, Segretain, \& M\'era \cite{csm}).  
There is a new constraint 
{}from Flynn, Gould, \& Bahcall (\cite{fgb}), which 
requires halo white dwarfs to have an
luminosity $M_V \ga 18.4.$ for $V-I > 1.8$.

The white dwarf luminosity function in our model 
is given in
Figure \ref{fig:wdlf}.  The two peaks
corresponding to the two bursts of star formation.
The strongest constraint comes from the more recent
peak produced 
at the epoch of
halo collapse.  These white dwarfs 
have cooled to present luminosities of
$\sim 10^{-5.5} L_\odot$.  This would correspond to a
bolometric magnitude of 18.4.  However, the bolometric
correction (Liebert, Dahn, \& Monet \cite{ldm}), 
would probably reduce the visible luminosity to 
$M_V > 20$, well below 
the Flynn, Gould, \& Bahcall limit (similar to the findings
of Kawaler \cite{kaw}).  
The low luminosity of the white dwarfs derives in part from their very old
age, but also because the population comes from a shifted IMF.
Since the progenitors are more massive stars, they died
sooner and so the remnants have had longer to cool.
Indeed, our adopted IMF centroid of $2.3 \msol$ is that of
Adams \& Laughlin (\cite{al}), who chose it specifically
to obey the luminosity function constraints in the disk.

\subsection{Parameter Sensitivity}
\label{sec:param}

Our results depend sensitively on several of the 
model parameters.  
Parameter values for our best-fit 
model are summarized in column 3 of Table 1.
Among the most important parameters
are those that determine the shape of the IMF.
We find a good fit for a centroid
(c.f.\ \S \ref{sec:imf}) $m_c = 2.3 \msol$, and a width 
$\sigma=1$.  Note that this centroid is that of 
Adams \& McLaughlin (\cite{al}).
However, they advocated a tighter width to avoid metallicity
problems which we do not require in our model
because metals are efficiently ejected in the
hot wind.
Indeed, the width we use is essentially the
present value (Miller \& Scalo \cite{milscal}).

\begin{table*}[htb]
\label{tab:modpar}
\caption{Model Parameters Adopted}
\begin{tabular}{cccc}
\hline
\hline
Model
Feature & Parameter & Best-Fit & Max Remnant  \\
\hline
IMF & $m_c$ & 2.3 \msol & same  \\
    & $\sigma$ & 1.6 & same   \\
Star Formation & \asfr & 0.007 & 0.002 \\
               & \bsfr & 1 & 2.5 \\
Ejecta & \emix & 8 & 8 \\
       & \zloss & 1 & 1 \\
\hline
\hline
\end{tabular}
\end{table*}

Other IMF parameters are allowed, but to obey the
observational constraints, these
must be near the ones we have adopted.  
As discussed above (\S \ref{sec:imf}), the centroid must
be $\ga 1 \msol$ to avoid significant production of 
long lived, low mass stars still burning in the halo.
However, the centroid must not be too high to avoid
untenably large metal yields.  The width must also be
not so large that it allows too many low mass stars,
but not so narrow that metal yields are too small and
helium yields too large.  Thus, the need for a dark halo,
and for a reasonable nucleosynthesis, drive the IMF
parameters to the range of those we have chosen.

The other important parameters involve star formation
and ejecta.  The star formation parameters 
(eq.\ \ref{eq:sfr} and Table 1) are reasonable
and consistent with previous results (MS93),
derived from requiring consistency
among various comoschronometers.  
The quiescent term dominates, controlling the mass budget,
but the merger term is also important.
The ratio of the merger to quiescent contribution controls the
position of the peak in the halo metallicity distribution
(Fig.\ \ref{fig:hmdis}).

For the ejecta, the most important parameter is the 
degree of mixing \emix.  A higher value leads to more
dilution of the ejecta, which leads to lower metallicity
and temperature in the hot gas.  The observed 
lower temperatures
and metallicity of the hot gas in groups demands that 
$\emix > 1$.

\section{The Remnant Contribution to the Dark Halo Mass}
\label{sect:machofrac}

Given that (1) there is good evidence for dark matter in
the Galactic halo, and that (2) microlensing objects
are the first positively detected dark matter in the halo,
a key question arises:  how much of the dark halo is in
MACHOs?  In this context it
is important to note that our best model does not give
an all-remnant halo.
Such a model is attractive for its simplicity, 
but we find that our best fit prefers only 
$\sim 40 \%$ of the dark halo mass in remnants, with the balance 
in non-baryonic material.
The strongest constraint on the white dwarf halo fraction
is the halo metallicity distribution.
Increasing the white dwarf fraction requires increasing 
the star formation rate; this leads to halo metallicities
whose distribution peaks too high (\S \ref{sec:param}).  
The ``maximum remnant'' model
is that having the $2\sigma$ 
upper limit for the halo fraction
(as set by the halo globular cluster
metallicity distribution).
It's parameters are summarized in Table 1; in it
the halo remnant mass fraction
is 77\%.  This
upper limit can be raised
somewhat if we increase the stellar merger dispersal efficiency
$\kappa$ (see \S \ref{sec:cloud}).  While it is 
difficult to increase the halo
remnant fraction, it is easy to decrease it.  If the winds are
less efficient (e.g., if $\zloss < 1$), then more material is
recycled in the clouds.  Thus a lower star formation amplitude is
required and fewer stars are available to eject at early times.
We find a minimum halo remnant fraction, for $\zloss \la 0.08$, of 
$\sim 1 \%$.

Our findings are to
be compared to recent analysis of the MACHO data
(Alcock et al.\ \cite{macho96}).
The microlensing results are still very model-dependent, but
it is provocative that they find a best fit for
$\sim 50\%$ of the halo being made up of 
$0.5^{+0.3}_{-0.2} \msol$ objects.
The question of whether one can put stronger constraints on
the MACHO halo fraction in our model leads to the
issues raised in the next section.

\subsection{X-ray Groups:  a Key Constraint}
\label{sec:xray}

As indicated in \S \ref{sec:data}, there is currently
large uncertainty regarding
X-ray observations of spiral dominated groups.
However, while the observations are ambiguous, what 
seems more
clear is that spiral-rich groups
do not contain hot gas with
$T \sim 1$ keV.  That is, these systems do not evidence
the same kind of diffuse emission as groups with
early-type galaxies.  If hot gas exists
in these groups, it must have
either low mass 
($\la (1-3) \times 10^{10} \msol$), 
or a low temperature, near or below the ROSAT
threshold $T_{\rm th} \la 0.3$ keV.
Indeed, Mulchaey, Mushotzky, Burstein, \& 
Davis (\cite{mmbd}) suggest that these systems do have cool
($T \la 0.3$ keV) gas, and argue that 
this may have been detected as high-ionization
quasar absorption lines.
Furthermore, given the morphological similarity 
of the bulge and halo with early-type galaxies,
it would be surprising if {\it no} hot gas were
found in these systems.  

At any rate, the observations seem to
rule out that the gas is
both massive and hot in spiral-rich groups.  
However, in our model
the temperature and the gas mass are related, as follows. 

After the processing of one stellar generation, 
the ejecta to remnant ratio is $\ret/(1-\ret)$.  Mixing with
unprocessed cold material gives
a total hot gas to remnant ratio $(1+\emix)\ret/(1-\ret)$.
In our model, $\ret = 0.38$, so $M_{\rm gas} > M_{\rm rem}$ for
$\emix > 1.7$ (which holds in our case).
This already means that 
the mass of ejected gas must be larger
than the halo baryon mass
(assuming that most hot gas escapes).  
Consequently, if the limits on gas in spiral-dominated  groups apply--i.e., if
the gas were to have $T \ga 0.3$ keV--then the mass in halo
white dwarfs must be less than the X-ray
limits on the gas mass,
$\sim 10^{10} \msol$.  This would be only a small component
of the halo.  

To allow for a significant gas mass, therefore, one
must demand that it avoid the ROSAT limits because it is cool:
$T \la 0.3$ keV.  However, the gas temperature is related to the
gas-to-remnant ratio, as we now show.  The total energy
in the ejecta is $\ret \esn M_{\rm rem}$.  This implies that
the energy per unit mass of ejected gas is 
$\epsilon_{\rm gas} = 
  \ret \esn M_{\rm rem}/M_{\rm gas} = \esn (1-\ret)/(1+\emix)$.
Finally, using the scaling 
$T_{\rm SN} = (2/3) \ \mu m_p \esn = 2$ keV, we have
$T_{\rm gas} = 2 \ {\rm keV} (1-\ret)/(1+\emix)$.  That is, the gas
temperature diminishes with \emix.  But
we already have seen
that having a significant fraction of the halo in remnants,
we need the concomitant halo gas to be cool.  Using the
Ponman et al.\ (\cite{pbeb}) value
$T_{\rm gas} \simeq 0.3$ keV, and with $\ret = 0.39$, 
one finds $\emix \simeq 3$.
In our detailed model which includes reprocessing, 
we find a larger value, $\eta = 8$.
Thus, cooler gas results if $\emix > 1$, but only
if the bulk of the baryons
are in the gas.

\subsection{Cosmological Baryon Fraction}

In the preceding section we see that our best-fit 
model finds that a large fraction of baryons, initially
in the group, are ejected as an intergalactic medium.
This can be reconciled with the 
total cosmological baryonic budget, as follows.
In our scenario, the cosmic baryonic inventory is
$\Omega_B = \Omega_{\rm disk+bulge} + 
\Omega_{\rm halo}^{\rm rem} + \Omega_{\rm LG}^{\rm gas} + 
\Omega_{\rm IGM}^{\rm gas} 
\simeq \Omega_{\rm halo}^{\rm rem} + \Omega_{\rm LG}^{\rm gas} + 
\Omega_{\rm IGM}^{\rm gas}$
(writing the universal density in component $i$ 
in units of the critical density:  $\Omega_i \equiv \rho_i/\rho_{\rm crit}$,
where $\rho_{\rm crit} =  3 H_0^2/8\pi G$).
If most baryons reside in the intragroup
and intergalactic diffuse gas, we therefore
require $\Omega_{\rm LG}^{\rm gas} + 
\Omega_{\rm IGM}^{\rm gas} \simeq \Omega_B \gg \Omega_{\rm halo}^{\rm rem}$.

Observationally, galactic rotation curves imply that
galaxy halos have 
$\Omega_{\rm halo} \sim 0.02 h (R_{\rm halo}/50 \ {\rm kpc})$,
where $R_{\rm halo}$ is the (unknown) radius of the dark halo
(e.g., Peebles \cite{peebles}).
On the other hand, primordial nucleosynthesis calculations give
$\Omega_B = (0.015 \pm 0.005) h^{-2}$ 
(Copi, Schramm, \& Turner \cite{cst}; Fields et al.\ \cite{fkot}).
Taking the ratio, we have 
$\Omega_{\rm halo}/\Omega_B \sim 1.3 h^3$. 
Consequently, for $h \la 0.9$, $\Omega_{\rm halo} < \Omega_B$;
that is, for reasonable values of $h$, some baryons
are likely to be nongalactic.  Indeed, for a low Hubble constant
($h = 0.5$), the {\it bulk} of the baryons are nongalactic.
Furthermore, these relations assume that the halos are entirely baryonic
({\it not} the case in our best model).  Any nonbaryonic component in
galaxy halos only strengthens the argument.  

Thus, we see that the X-ray observations of groups can
provide strong constraints on remnants in the halo.
Of course, this assumes that the Local Group is like other 
poor groups.
If so, and if the X-ray data is reliable, 
then these data may provide a key constraint.  
As we have noted, if there is only a small X-ray 
gas mass in the Local Group,
then there can be few remnants in the halo.  
Taking the more optimistic view, if the Ponman et al.\ (\cite{pbeb})
result is correct, then there could be 
a large amount of (as yet unobserved) hot gas in the Local Group, 
as required in our model.

\section{Conclusions} 
\label{sec:imp}

We have shown that,
without violating
constraints posed by 
luminosity and nucleosynthesis considerations,
one may construct 
a plausible model 
in which the dark halo
of the Galaxy 
contains a significant fraction of white dwarfs.
These may have already
been detected in halo microlensing events towards the LMC,
and might also be detected via their luminosity function.
The same bursts of star formation which produced the 
white dwarfs also led to hot, metal-rich intergalactic gas,
some of which may still reside in the 
Local Group.
This hot gas could be detectable via its 
X-rays, and by distortions
in the cosmic microwave background radiation (Suto et al.\ \cite{smio}).

Thus, the predictions of the model are testable.  If the halo is comprised
of white dwarfs then there must be a background of hot, X-ray 
emitting gas in the Local Group.  Conversely, if there is 
metal-rich hot gas in the Local Group, then 
a significant fraction of
the halo mass must be in remnants.
Clearly, further searches for both of these are warranted.

Furthermore, if our galaxy formation scheme is indeed universal,
then hot gas production and ejection should be a ubiquitous 
aspect of halo formation.  
Consequently, X-ray observations of other systems could
provide
a key constraint on our model.  In particular, 
our model can be directly tested by
observations which can unambiguously 
confirm or deny the presence of
hot gas in other spiral-dominated groups.
Also, if white dwarfs are ubiquitous in galactic
halos, then
they may lead to detectable infrared profiles in edge-on galaxies,
which may already have been observed
(Barnaby \& Thronson \cite{bt94};
Sackett, Morrison, Harding, \& Boroson \cite{smhb};
Lequeux, Fort, Dantel-Fort, Cuillandre, \& Mellier \cite{lfdcm};
Lehnert \& Heckman \cite{lh}).  
Finally,  even if our scenario
turns out not to be applicable to spiral-dominated groups,
it remains that ellipticals must eject gas in strong winds.
Thus, our model may still be valid for clusters, which
are elliptical-dominated.  This will be explored in a subsequent work.

We note as well that
in our scenario, just as there is typically
a large 
outflow from the halo, there is also a strong 
evaporative wind that 
ejects material from the Local Group.
As a result, most baryons eventually reside 
in hot, intergalactic (as
opposed to intragroup) gas.  If this gas stays hot, 
it could perhaps be the 
ionized 
intergalactic (as suggested by Gunn-Peterson limits on the neutral 
intergalactic medium).  
If it does cool, it presents serious problems, as 
it would lead to prodigious but unobserved absorption of extragalactic
radiation.

Finally, we reiterate that stellar remnants and their 
associated hot
ejecta are conservative
candidates for both the halo microlensing objects and for
the baryonic dark matter.  If these can
be ruled out, then we are forced to conclude that 
the microlensing objects and the dark baryons are something 
stranger still.

\acknowledgments
\nobreak
We are pleased to acknowledge useful discussions with 
C. Alcock, D. Bennett, E. Gates, K. Jedamzik, 
S. Shore, and M. Turner.
We are grateful to R. Mushotzky and J. Mulchaey for 
may useful discussions
of X-ray observations, and we especially thank R. Pildis
for pointing out the limits on spiral-dominated groups.
Finally, we are particularly indebted to St\'ephane Charlot
for advice and assistance on an early version of this work.
Work at Notre Dame is supported by DoE
Nuclear Theory grant number DE-FG02-95ER40934.
Work at the University of Chicago 
is supported by NSF grant AST 90-22629, DOE grant DEF02-91-ER40606, 
and NASA grant 1231, and by NASA through grant NAGW 2381 at Fermilab.

\newpage

\centerline{FIGURE CAPTIONS}

\begin{enumerate}

\item {\label{fig:diag}
Schematic diagram of model features.}

\item {\label{fig:sfr}
The total galactic star formation rate $\psi = n_{\rm c} \psi_{cloud}$ 
(solid curve) and number of clouds $n_{\rm c}$ 
(dashed curve) as a function of time.
Two bursts of star formation are evident, 
one shortly after decoupling when the density is high,
and one during halo collapse (5 Gyr).}

\item {\label{fig:masses}
Mass evolution for (a) the halo, (b) the local Group.}

\item {\label{fig:mets}
Metallicity evolution for the halo and Local Group.}

\item {\label{fig:hmdis}
Halo metallicity distribution.  The points
are for globular clusters (Pagel \protect\cite{pagel});
only the low metallicity members ($[Z] < -1$)
are included in the analysis.  
The theory (histogram) is binned in the
same way as the data, and normalized
to minimize $\chi^2$.}

\item {\label{fig:wdlf}
The halo white dwarf luminosity function,
with $L$ in units of $L_\odot$. 
Only results for halo white dwarfs are shown.
The two peaks correspond to the
two bursts of star formation; the more luminous peak
is the more recent burst at $t=5$ Gyr.  This
second peak has $M_V \ga 20$, well below current 
observational limits.}

\end{enumerate}

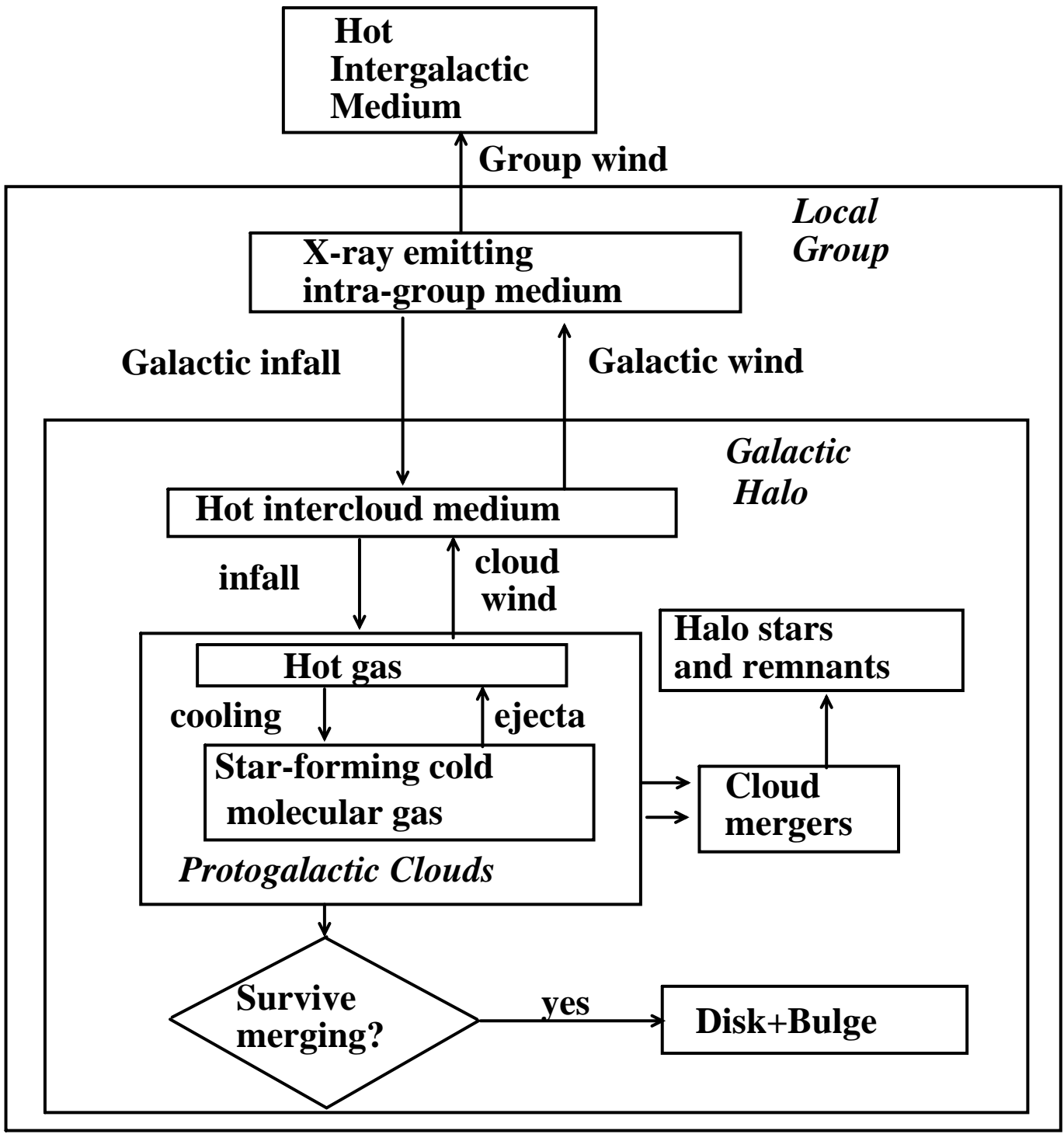
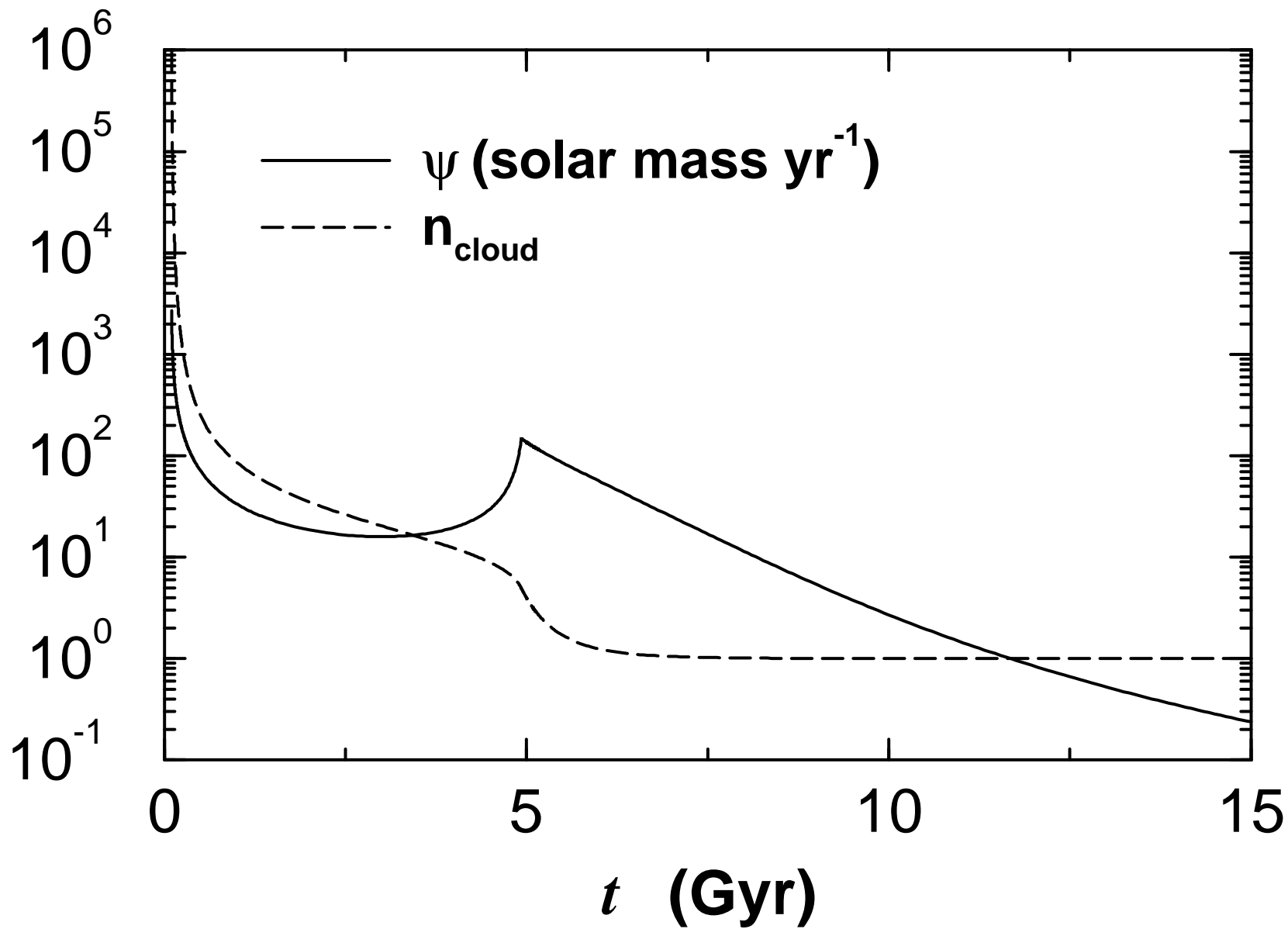
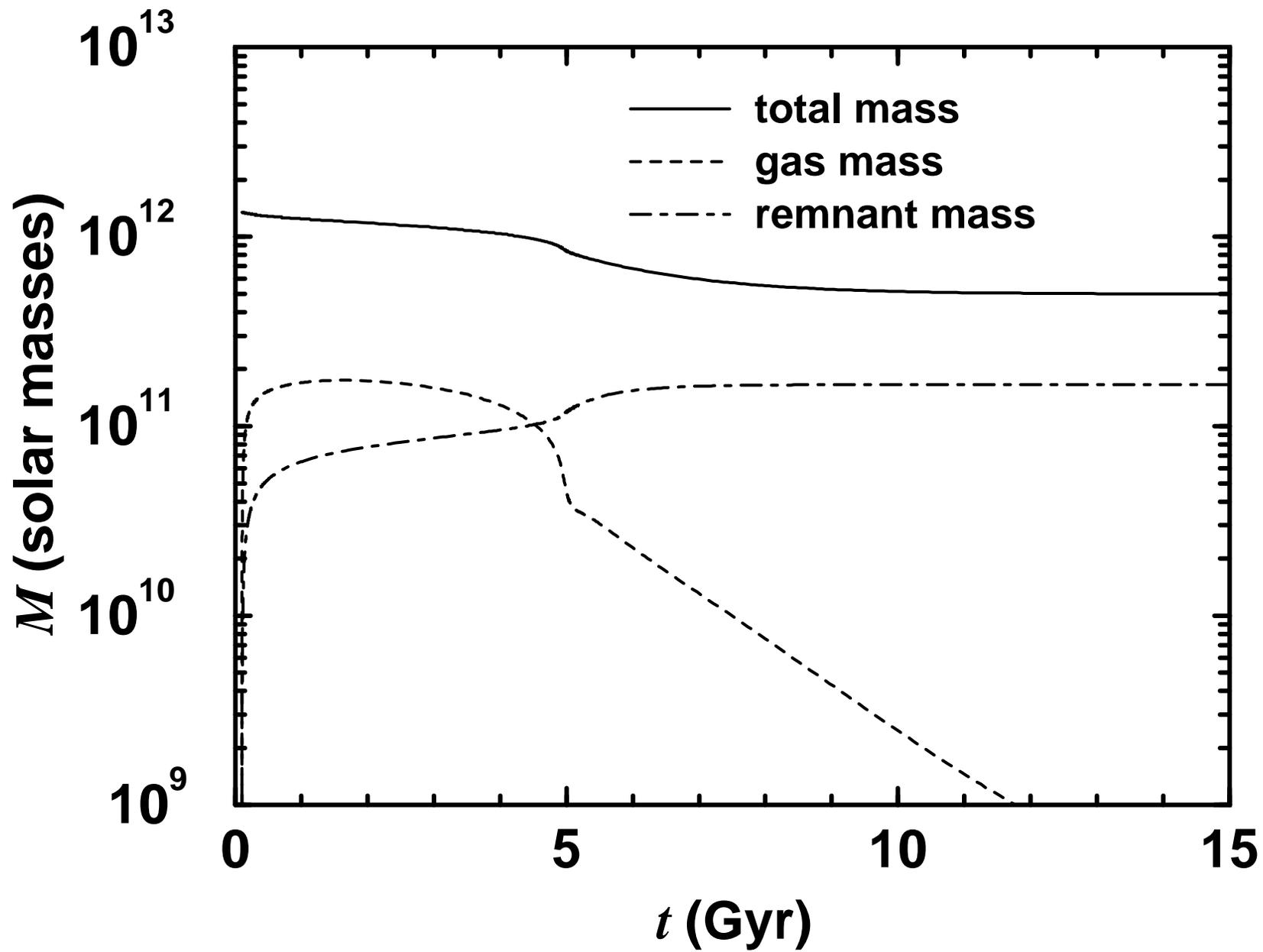
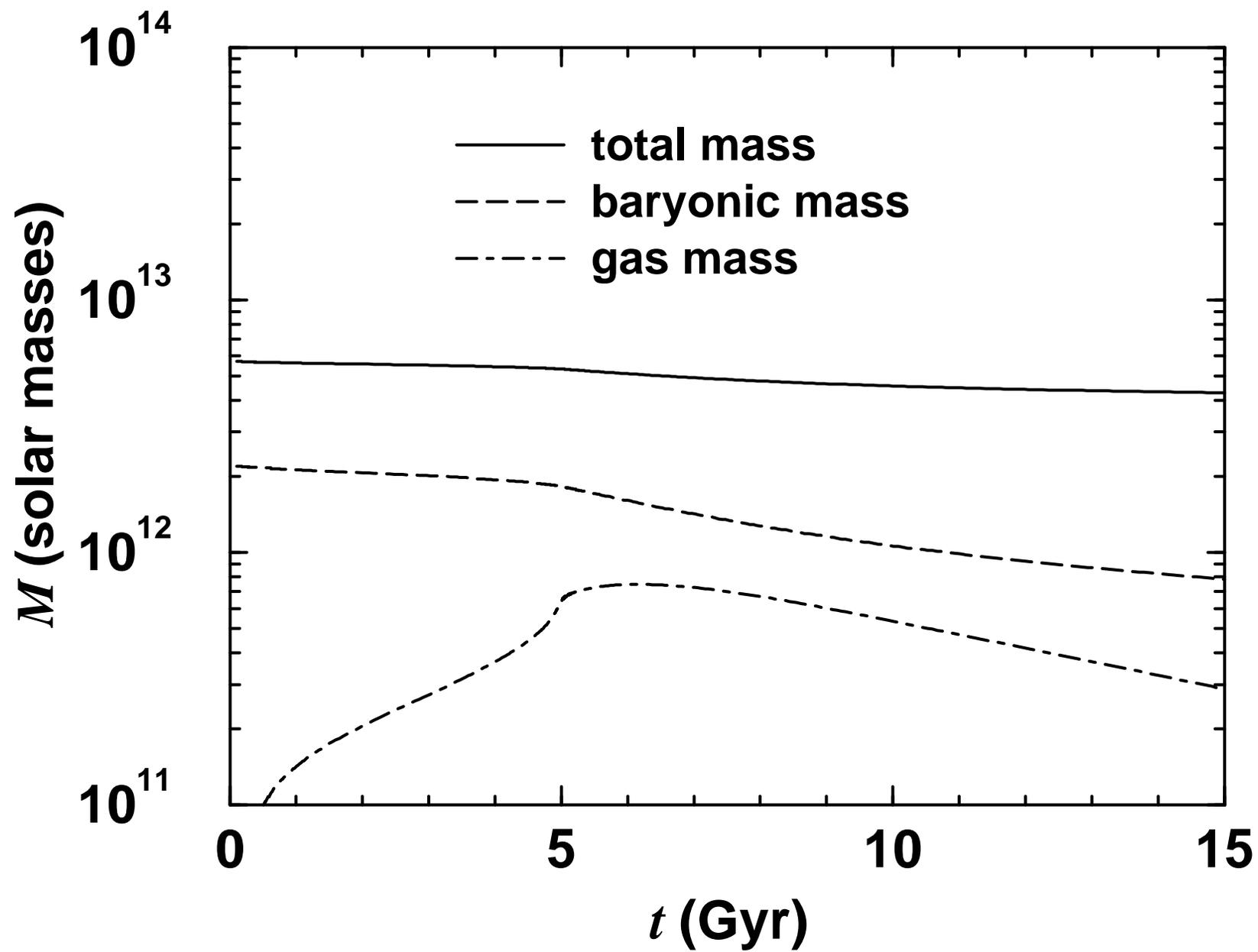
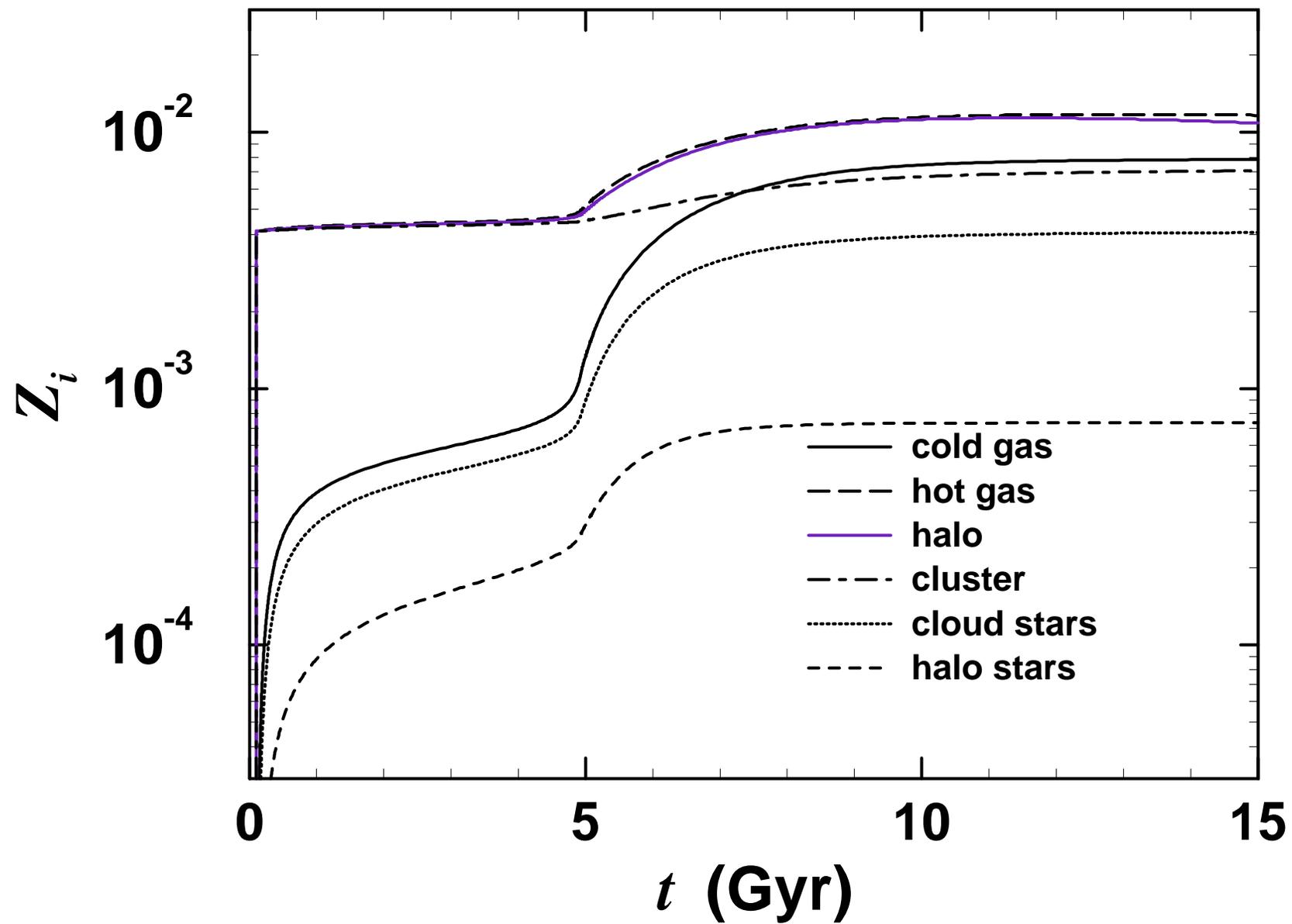
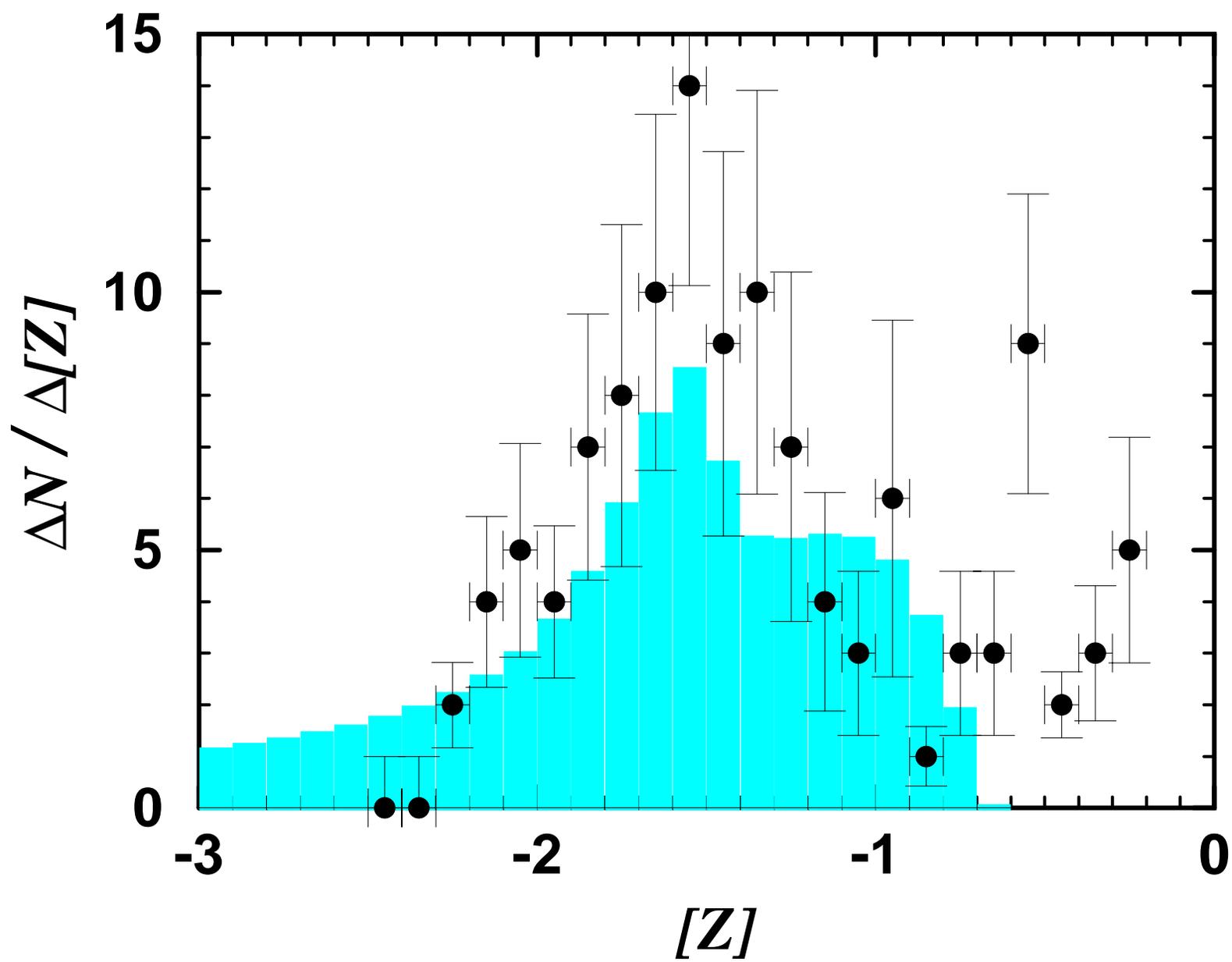
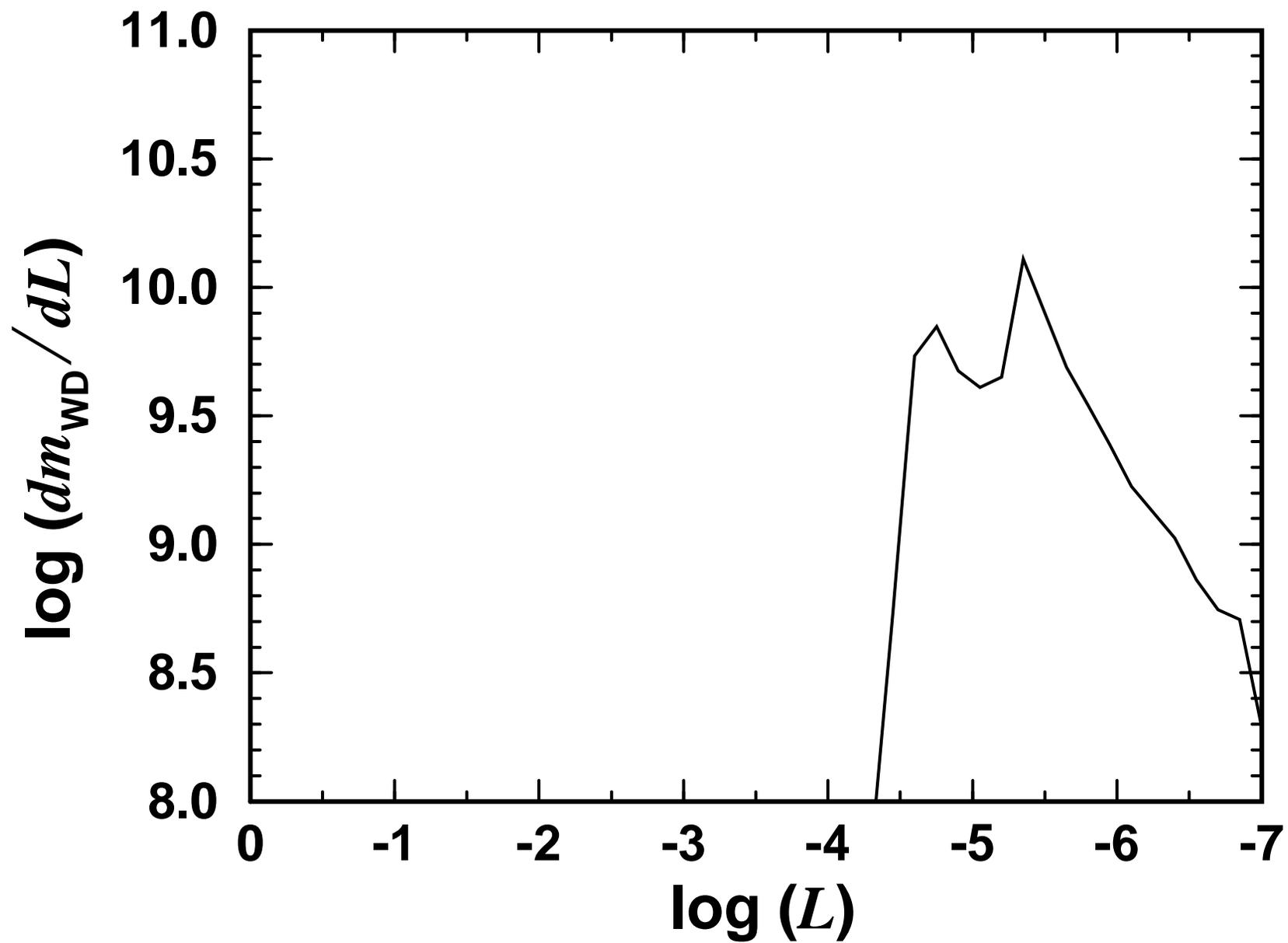
\end{document}